\documentclass[twocolumn,pre,showpacs,superscriptaddress]{revtex4-1}
\usepackage[left=2.5cm,right=2.5cm,top=2cm,bottom=3cm,includeheadfoot]{geometry}
\usepackage{graphicx}
\usepackage{rotating}
\usepackage{amsmath}
\usepackage{amsfonts}
\usepackage{amssymb}
\usepackage{enumerate}
\usepackage{longtable}
\usepackage{units}
\usepackage{hyperref} 
\usepackage[T1]{fontenc}
\usepackage[latin1]{inputenc}
\usepackage{graphicx}
\usepackage{capt-of}
\usepackage{bbm}
\usepackage{dcolumn}% Align table columns on decimal point
\usepackage{bm}
\usepackage{color}

\begin{document}
\newcommand{\cred}{\color{red}}
\title{Optimizing linked cluster expansions by white graphs}

 \author{K. Coester}
 \affiliation{Lehrstuhl f\"ur Theoretische Physik I, Otto-Hahn-Str.~4, TU Dortmund, D-44221 Dortmund, Germany}
 
 \author{K.P. Schmidt}
 \affiliation{Lehrstuhl f\"ur Theoretische Physik I, Otto-Hahn-Str.~4, TU Dortmund, D-44221 Dortmund, Germany}

\date{\rm\today}

\begin{abstract}
We introduce a {\it white graph expansion} for the method of perturbative continuous unitary transformations when implemented
 as a linked cluster expansion. The essential idea behind an expansion in white graphs is to perform an optimized bookkeeping 
during the calculation by exploiting the model-independent effective Hamiltonian in second quantization and 
the associated inherent cluster additivity. This appoach is shown to be especially well suited for microscopic models 
with many coupling constants, since the total number of relevant graphs is drastically reduced. 
The white graph expansion is exemplified for a two-dimensional quantum spin model of coupled two-leg XXZ ladders.
\end{abstract}


\maketitle

\section{Introduction}
\label{Sect:Intro}

High-order series expansions are an important tool in statistical physics. Typically, the linked cluster theorem is used to determine the correct expression of physical quantities up to high orders in perturbation by performing calculations on finite linked clusters. Historically, such {\it linked cluster expansions} date back to the 1950's and 1960's \cite{Rushbrooke1955,Rushbrooke1964,Baker1966,Baker1967}, where high-temperature series for extensive thermodynamic quantities have been determined. Similarly, extensive zero-temperature ground-state properties like the ground-state energy or susceptibilities can be calculated via linked cluster expansions as first proposed by Nickel in 1980 and implemented by Marland in 1981 \cite{Marland1981}. Further progress has been made steadily over the years \cite{Irving1984,He1990,Zheng1991,Hamer1994}.

For non-extensive quantities like excitation energies, however, linked cluster expansions are more complicated and for a long time the calculation of an energy gap was only achieved by a deficient scheme which also needed \textit{unlinked} clusters \cite{He1990}. It took until 1996 when Gelfand set up a true linked cluster expansion for a one-particle dispersion \cite{Gelfand96}. Gelfand realized that quantum fluctuations of the vacuum have to be subtracted from one-particle hopping amplitudes to perform proper linked cluster expansions. Yet, unrealized at that time, this approach in terms of similarity transformations on graphs violates the cluster additivity and is therefore inapplicable when the ground state and the targeted excitation-subspace are characterized by identical quantum numbers. In 2000. it has been shown that the use of orthogonal transformations on graphs restores the cluster addiditivity and therefore allows for a consistent calculation of many-particle excitation energies \cite{Trebst2000,Zheng2001,Oitmaa2006}.

At the same time, an alternative route to linked-cluster expansions has been established by the method of perturbative continuous unitary transformations (pCUTs) \cite{Knetter1999} allowing for the calculation of high-order series expansions for many-particle excitation energies as well as spectral densities \cite{cc1}. In contrast to the other approaches mentioned above, in pCUTs a quasi-particle conserving effective Hamiltonian in second quantization is derived model-independently, with the constrained that the unperturbed part of the Hamiltonian has an equidistant spectrum and is bounded from below. Interestingly, this method fulfills the cluster additivity by construction. In recent years, pCUTs were indeed used as linked-cluster expansions, i.e.~a full graph decomposition has been implemented to calculate relevant matrix elements. Important examples are the derivation of effective low-energy spin models for the Hubbard model on the triangular and honeycomb lattice \cite{Yang2010, Yang2012}, the calculation of the one-magnon gap for the fully-frustrated transverse field Ising model on the triangular and kagome lattice \cite{Powalski2013}, the treatment of topological phase transitions of perturbed non-Abelian string-net models on the honeycomb lattice \cite{Schulz2013,Schulz2014}, or the determination of the one-triplon dispersion of coupled Heisenberg dimers on hypercubic lattices for arbitrary dimension \cite{Joshi2015}.

Overall, linked cluster expansions constitute an efficient tool with a vast variety of applications. At this point, let us also mention the recently developed non-perturbative variants of linked cluster expansions combining a graph decomposition with exact diagonalization \cite{Rigol2006,Kallin2013,Rigol2014_1,Tang2015}, with continuous unitary transformations \cite{Yang2010,Yang2012,Ixert2014,Coester2014} or density matrix renormalization group \cite{Stoudenmire2014}. 

The numerical power of all (perturbative) linked-cluster approaches, especially in more than one dimension, relies on a full graph decomposition. To this end the lattice is devided into small subclusters on which the actual calculations are carried out. Afterwards, results in the thermodynamic limit are obtained by embedding properly the finite-cluster results into the infinite system. If all sites of two different subclusters are coupled identically, these clusters are indistinguishable for the Hamiltonian and are called topologically equivalent. Therefore, the calculation can be restricted to topologically distinct clusters yielding a highly efficient approach.

However, the number of topologically distinct graphs grows exponentially with the order of perturbation as well as with the total number of different coupling constants. Consequently, it becomes very hard to reach sufficiently high orders for problems with several expansion parameters. This is especially relevant for the comparison with experimental data where typically different coupling strengths are important and have to be determined. 

In this work we introduce a new kind of graph expansion, a so-called {\it white graph expansion}, which overcomes the latter limitation to a great extent. The white graph expansion benefits directly from the underlying framework provided by the effective pCUT Hamiltonian in second quantization. 

The paper is organized as follows. We start by giving the set up of systems we focus on in this paper in Sect.~\ref{Sect:set_up}. In Sect.~\ref{Sect:pCUT} we describe the pCUT method and its implementation as a linked cluster expansion. The white graph expansion is then introduced in Sect.~\ref{Sect:White_Graph_Expansion} and it is applied to coupled two-leg XXZ ladders in Sect.~\ref{Sect:XXZ_Ladders}. Finally, we give conclusions in Sect.~\ref{Sect:Conclusions}. 

%
%
%%%%%%%%%%%%%%%%%%%%%%%
%%%%%%%%%%%%%%%%%%%%%%%
\section{Set up}
\label{Sect:set_up}
%%%%%%%%%%%%%%%%%%%%%%%
%%%%%%%%%%%%%%%%%%%%%%%
%
%
We consider a generic quantum lattice Hamiltonian $\cal{H}$ at zero temperature. By decomposing the original lattice into a superlattice of supersites, one can always rewrite \emph{exactly} ${\cal H}$ as 
\begin{equation}
\label{Eq:Hami}
{\cal H}={\cal H}_0+\sum_{j=1}^{N_\lambda}\lambda_j {\cal V}^{(j)} \quad .
\end{equation}
Here a supersite might be a spin, two linked spins like a dimer, or any other finite set of linked sites which can be easily diagonalized and which has an equidistant spectrum bounded from below. 

The unperturbed part of $\mathcal{H}$ is diagonal in supersites $i$ of the lattice and can be written as
\begin{eqnarray}\label{h_0_q}
\mathcal{H}_0 &=& E_0+\sum_{i,\alpha} \hat{f}^\dagger_{i,\alpha}\hat{f}^{\phantom{\dagger}}_{i,\alpha}\nonumber\\
              &=& E_0+\mathcal{Q} \quad ,
\end{eqnarray}
where $E_0$ denotes a constant and the sum over $\alpha$ runs over all excited local degrees of freedom. The unperturbed ground state $|{\rm ref}\rangle$ is interpreted as the vacuum and is given as the product state $|{\rm ref}\rangle\equiv |0\rangle \ldots |0\rangle $ with $|0\rangle$ being the lowest state of a supersite. Accordingly, $\hat{f}^\dagger_{i,\alpha}|{\rm ref}\rangle$ creates a local excitation of type $\alpha$ on supersite $i$. In the following we assume the local spectrum to be equidistant. Consequently, it is always possible to introduce the counting operator \mbox{$\mathcal{Q}\equiv\sum_i\hat{n}_i\equiv \sum_{i,\alpha} \hat{f}^\dagger_{i,\alpha}\hat{f}^{\phantom{\dagger}}_{i,\alpha}$}.
 
Supersites interact via the perturbation \mbox{$\mathcal{V}\equiv\sum_j\lambda_j\mathcal{V}^{(j)}$}. The sum over $j$ runs over all $N_\lambda$ perturbation parameters $\lambda_j$. The different operators $\mathcal{V}^{(j)}$ build the bonds of the lattice, so that one can assign a different "color" for each perturbation parameter $\lambda_j$. Here we restrict the discussion to Hamiltonians where the perturbation ${\cal V}$ couples two supersites. A generalization to perturbations which couple an arbitrary number of supersites is straightforward.
 
As a consequence of Eq.~\eqref{h_0_q}, one can rewrite Eq.~\ref{Eq:Hami} as
\begin{equation}
\label{Eq:Hami_final}
{\cal H}={\cal H}_0+ \sum_{n=-N}^N T_n \quad ,
\end{equation}
so that $[\mathcal{Q},T_n]=nT_n$. Physically, the operator $T_n \equiv\sum_j \lambda_j T^{(j)}_n$ corresponds to all operators where the change of energy quanta with respect to $\mathcal{H}_0$ is exactly $n$. The maximal (finite) change in energy quanta is called $\pm N$. 

%
%
%%%%%%%%%%%%%%%%%%%%%%%
%%%%%%%%%%%%%%%%%%%%%%%
\section{Perturbative continuous unitary transformations}
\label{Sect:pCUT}
%%%%%%%%%%%%%%%%%%%%%%%
%%%%%%%%%%%%%%%%%%%%%%%
%
%
The pCUT method is an efficient tool for the calculation of high-order series expansions for Hamiltonians of the form Eq.~\eqref{Eq:Hami_final} \cite{Knetter1999,cc1}. The application consists essentially of two steps: i) the perturbative order-by-order solution of the so-called flow equation yielding an effective Hamiltonian and effective observables and ii) a non-trivial extraction of the effective low-energy physics. 

The first step depends only on the structure of the perturbation, namely, how many particles at most are created (annihilated) by the perturbation. In this sense, the obtained solution is model independent. However, this generality comes at the price of a complex model dependend extraction process because the effective operators are not normal-ordered, i.e.~the desired matrix elements have to be extracted in a separate (second) step. The real challenge lies in this second step which represents the model-dependend part and therefore typically is the bottle-neck of the calculation. The extraction process is based on the linked-cluster theorem which allows the determination of matrix elements in the thermodynamic limit by applying the effective operators to \textit{finite} clusters. At the end of this process a normal-ordered low-energy Hamiltonian is obtained.

Essentially, two different schemes for the last step are possible. First, in the standard scheme the calculation is carried out on a single cluster \cite{Knetter1999,Knetter2000,Schmidt2003,Dorier2008,Vidal2009} which is sufficiently large to contain all processes of a desired perturbative order. The implementation of this
 approach is straightforward and details of the operator structure are irrelevant for the general scheme. However, in two or more dimensions, this is typically unfavourable because memory problems arise and a full cluster decomposition is the method of choice \cite{Yang2010,Powalski2013,Schulz2013,Schulz2014}. 
 
Let us also mention that a hybrid, a decomposition into rectangular clusters, has been also succesfully introduced recently for the pCUT method \cite{Dusuel2010}. A rectangular cluster expansion seems to be especially well suited for non-perturbative extensions of linked cluster expansions, since a typical length scale can be assigned to each rectangular graph, which is much less trivial when performing a full graph decomposition \cite{Kallin2013,Coester2014}.

In the following we start by briefly introducing the method of continuous unitary transformations and its perturbative variant, the pCUT method. Afterwards, we formulate the pCUT method as a linked cluster expansion giving details to graph generation and graph embedding.

\subsection{Perturbative solution of the flow equation}
The objective of continuous unitary transformations (CUTs) is to transform the Hamiltonian into an optimized basis representation \cite{Wegner1994,Wilson1994}. The Hamiltonian is then considered as a continuous function $\mathcal{H}(\ell)$ of the flow parameter $\ell$ with \mbox{$\mathcal{H}(\ell=0)=\mathcal{H}$} as the starting Hamiltonian and \mbox{$\mathcal{H}(\ell=\infty)=\mathcal{H}_\text{eff}$}  as the effective Hamiltonian. Introducing the antihermitian generator $\eta(\ell)$ of the CUT, one gets the flow equation
\begin{equation}
\label{Eq:flow}
\frac{\text{d}\mathcal{H}(\ell)}{\text{d}\ell}=\left[\eta(\ell), \mathcal{H}(\ell)\right] \; . 
\end{equation}

Here we choose the quasi-particle generator~\cite{Knetter1999,Mielke1998,cc1}
\begin{equation}
\label{Eq:generator}
\eta_{i,j}(\ell)=\text{sgn}(q_i-q_j) \mathcal{H}_{i,j}(\ell)
\end{equation}
in an eigenbasis $|i\rangle$ of the counting operator $\mathcal{Q}$, i.e.~\mbox{$\mathcal{Q}|i\rangle=q_i |i\rangle$}, as introduced in Eq.~\eqref{h_0_q}. 

The commutator in the flow equation Eq.~\eqref{Eq:flow} leads typically to an infinite number of terms, so that a truncation must be performed. In the pCUT method, the truncation is done in a perturbative manner and, accordingly, the flow equation is solved via the perturbative ansatz
\begin{eqnarray}
\label{Eq:ansatz}
\mathcal{H}(\{\lambda_j\};\ell)&=&\mathcal{H}_0+\\
        && \sum_{\sum_j k_j=k}^{\infty} \lambda_1^{k_1}\ldots\lambda_{N_\lambda}^{k_{N_\lambda}}\sum_{|\underline{m}|=k}F(\ell;\underline{m})T(\underline{m}),\nonumber
\end{eqnarray}
implying, according to Eq.~\eqref{Eq:generator},
\begin{eqnarray}
\eta^{\mathcal{Q}}(\{\lambda_j\};\ell)&=& \sum_{\sum_j k_j=k}^{\infty} \lambda_1^{k_1}\ldots\lambda_{N_\lambda}^{k_{N_\lambda}} \sum_{|\underline{m}|=k} {\phantom{\quad\quad\quad\quad\quad .}}\nonumber\\
         && {\phantom{\quad\quad\quad .}}F(\ell;\underline{m})\,\text{sgn} \left({M(\underline{m})}\right) T(\underline{m}).
\end{eqnarray}
Here, the following notations have been introduced
\begin{align}
\underline{m}&=(m_1,m_2,\dots , m_k)\\
m_i &\in \{0,\pm 1, \pm 2, \dots , \pm N\}\\
|\underline{m}|&=k\\
T(\underline{m})&=T_{m_1}T_{m_2}T_{m_3}\dots T_{m_k}\\
M(\underline{m})&=\sum_{i=1}^k m_i.
\end{align}
Ansatz \eqref{Eq:ansatz} leads to recursive differential equations for the real functions $F(l;\underline{m})$ with the initial conditions $F(0,\underline{m})=1$ for $|\underline{m}|=1$ and $F(0,\underline{m})=0$ for $|\underline{m}|>1$ recovering the initial Hamiltonian. Due to the structure of the recursive equations, they can be solved in principal analytically. The calculation of the analytic functions $F$ can be performed on a computer up to a certain order $n_\text{max}$ limited by the computation time and the memory usage only \cite{Knetter1999}. 

The solution leads to an effective Hamiltonian of the form
\begin{eqnarray}
\mathcal{H}_\text{eff}(\{\lambda_j\})&=&\mathcal{H}_0+ \sum_{\sum_j k_j=k}^{\infty} \lambda_1^{k_1}\ldots\lambda_{N_\lambda}^{k_{N_\lambda}}{\phantom{\quad\quad\quad\quad\quad .}}\nonumber\\
       &&{\phantom{\,\,\,\,\, .}}\sum_{|\underline{m}|=k\sum_i m_i=0}C(\underline{m})T(\underline{m}),\label{Heffpcut}
\end{eqnarray}
with $C(\underline{m})\in \mathbb{Q}$. The restriction $\sum_i m_i=0$ reflects the particle-conserving property $[\mathcal{H}_{\rm eff},\mathcal{Q}]=0$ of the final Hamiltonian. The summands can be viewed as virtual fluctuations of the new dressed particles defined by the effective Hamiltonian.

In a similar manner one can calculate other physical quantities with pCUT \cite{cc1}. To this end the same unitary transformation has to be applied to any observable $\mathcal{O}$ of interest
\begin{align}
\frac{\partial \mathcal{O}(\{\lambda_j\};\ell)}{\partial \ell}=[\eta^{\mathcal{Q}}(\{\lambda_j\};\ell),\mathcal{O}(\{\lambda_j\};\ell)].
\end{align}
Using the perturbative ansatz
\begin{eqnarray}
\mathcal{O}(\{\lambda_j\};\ell)&=&\sum_{\sum_j k_j=k}^{\infty} \lambda_1^{k_1}\ldots\lambda_{N_\lambda}^{k_{N_\lambda}}{\phantom{\quad\quad\quad\quad\quad\quad\quad\quad .}}\nonumber\\
&&{\phantom{\,\,\, .}}\sum_{i=1}^{k+1}\,\,\sum_{|\underline{m}|=k}G(\ell;\underline{m},i)\mathcal{O}(\underline{m};i)
\end{eqnarray}
with
\begin{align}
\mathcal{O}(\underline{m};i):=T_{m_1}\dots T_{m_{i-1}}\mathcal{O}T_{m_{i}}\dots T_{m_{k}}\quad,
\end{align}
one obtains recursive differential equations for the functions $G(\ell;\underline{m},i)$. The final result is given by
\begin{eqnarray}
\mathcal{O}_\text{eff}(\{\lambda_j\})&=&\sum_{\sum_j k_j=k}^{\infty} \lambda_1^{k_1}\ldots\lambda_{N_\lambda}^{k_{N_\lambda}}{\phantom{\quad\quad\quad\quad\quad\quad\quad\quad .}}\nonumber\\
 && {\phantom{\,\; .}}\sum_{i=1}^{k+1}\,\sum_{|\underline{m}|=k}\tilde{C}(\underline{m};i)\mathcal{O}(\underline{m};i)\label{Oeffpcut},
\end{eqnarray}
with $\tilde{C}(\underline{m},i)=G(\ell=\infty;\underline{m},i) \in \mathbb{Q}$. In contrast to the effective Hamiltonian, effective observables are \textit{not} quasi-particle conserving.

The coefficients $C(\underline{m})$ and $\tilde{C}(\underline{m};i)$ are independent of the model and can be straightforwardly applied to all the problems matching the requirements of the pCUT method. This generality is only possible at the expense of the second model-dependend extraction process described next for the effective Hamiltonian $H_{\rm eff}$. The treatment of effective observables can be done in complete analogy.

\subsection{pCUT as a linked cluster expansion}
\label{ssec:pCUT_LCE}

\begin{figure}
\begin{center}
\includegraphics*[width=0.95\columnwidth]{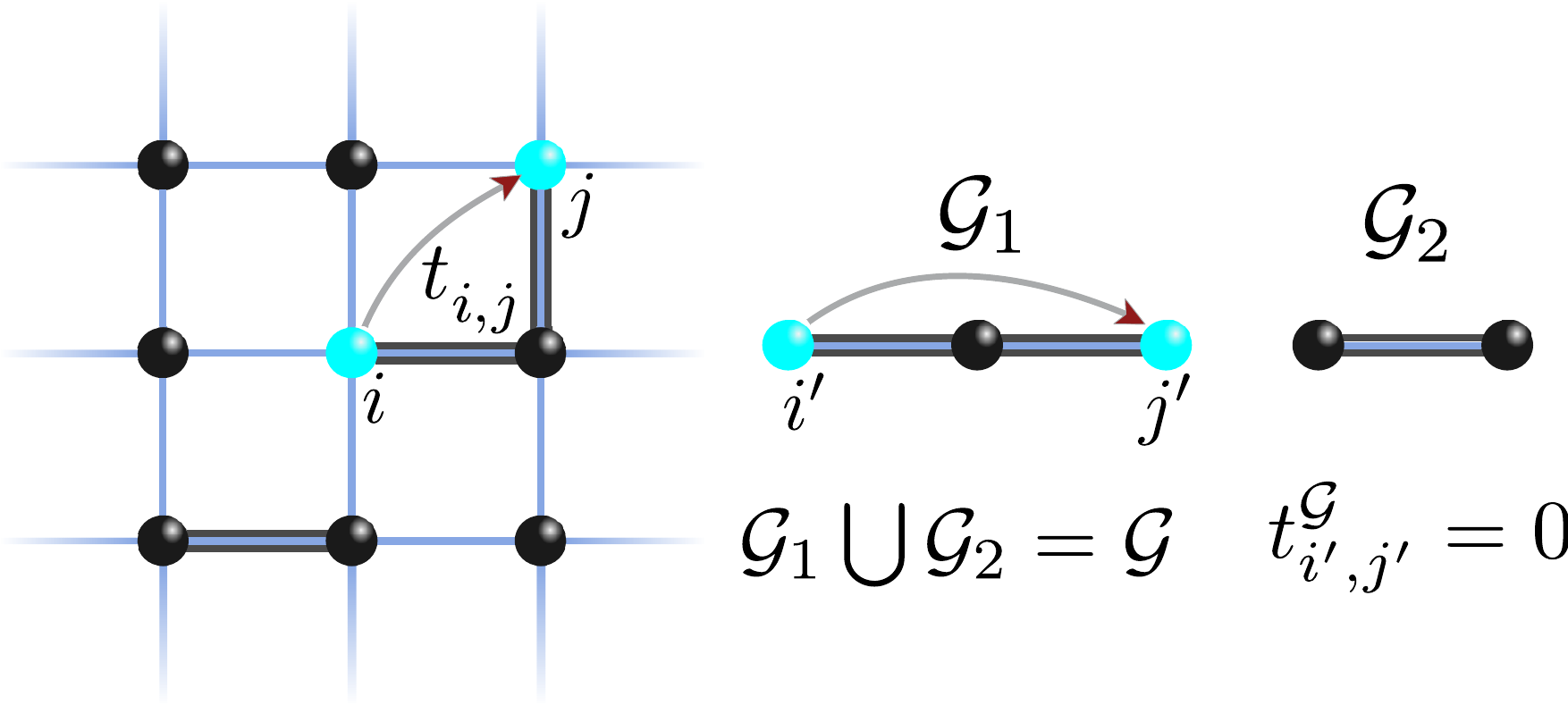}
\end{center}
\caption{All fluctuations involving the highlighted links are pooled together in this representation. The overall contribution of these fluctuations to the hopping element $t_{i,j}$ in the thermodynamic limit (left panel) can be identified with a reduced contribution of a disconnected cluster (right panel) vanishing due to the linked cluster theorem.} 
\label{fig:disconnected_fluctuations}
\end{figure}

Since the effective operators like $H_{\rm eff}$ are not normal-ordered, matrix elements defining the effective model are not directly accesible. In order to calculate the desired matrix elements, the effective Hamiltonian $H_{\text{eff}}$ can be applied to a finite cluster yielding the correct contributions in the thermodynamic limit for a given order and matrix element. This crucial property results from the linked cluster theorem which states that only linked processes have an overall contribution to cluster additive quantities.

As outlined in Sect.~\ref{Sect:set_up}, here we restrict the discussion to Hamiltonians where the perturbation couples always two supersites. The coupling between supersites is represented by a link in the cluster and the operator can be decomposed as
\begin{align}
T_n=\sum_l \tau_{n,l}\quad ,\label{tausumme}
\end{align}
where $\tau_{n,l}$ affects only the two supersites which are connected by the link $l$ in the effective lattice. We say, that the operator $\tau_{n,l}$ acts on the link $l$. The sum runs over all links of the lattice. Inserting this decomposition  into Eq.~\eqref{Heffpcut} yields
\begin{eqnarray}
\mathcal{H}_\text{eff}(\{\lambda_j\})&=&\mathcal{H}_0+\sum_{\sum_j k_j=k}^{\infty} \lambda_1^{k_1}\ldots\lambda_{N_\lambda}^{k_{N_\lambda}}\\
              &&\sum_{|\underline{m}|=k\sum_i m_i=0}\hspace*{-3mm}C(\underline{m})\sum_{l_1\dots l_k} \tau_{m_k,l_1}\dots \tau_{m_k,l_k}\,.\nonumber\label{full_decomposition}
\end{eqnarray}
In practice, however, the Hamiltonian is applied using Eq.~\eqref{Heffpcut}.
Each summand of order $k$ can be understood as a virtual fluctuation involving the links $l_1\dots l_k$. The links form a pattern which can be assigned to a cluster consisting of these links and adjacent sites. If the fluctuation pattern corresponds to a disconneted cluster, the overall contribution of all summands associated with this pattern annihilate each other and such fluctuation patterns can be discarded, see also Fig.~\ref{fig:disconnected_fluctuations}.

\begin{figure}
\begin{center}
\includegraphics*[width=0.95\columnwidth]{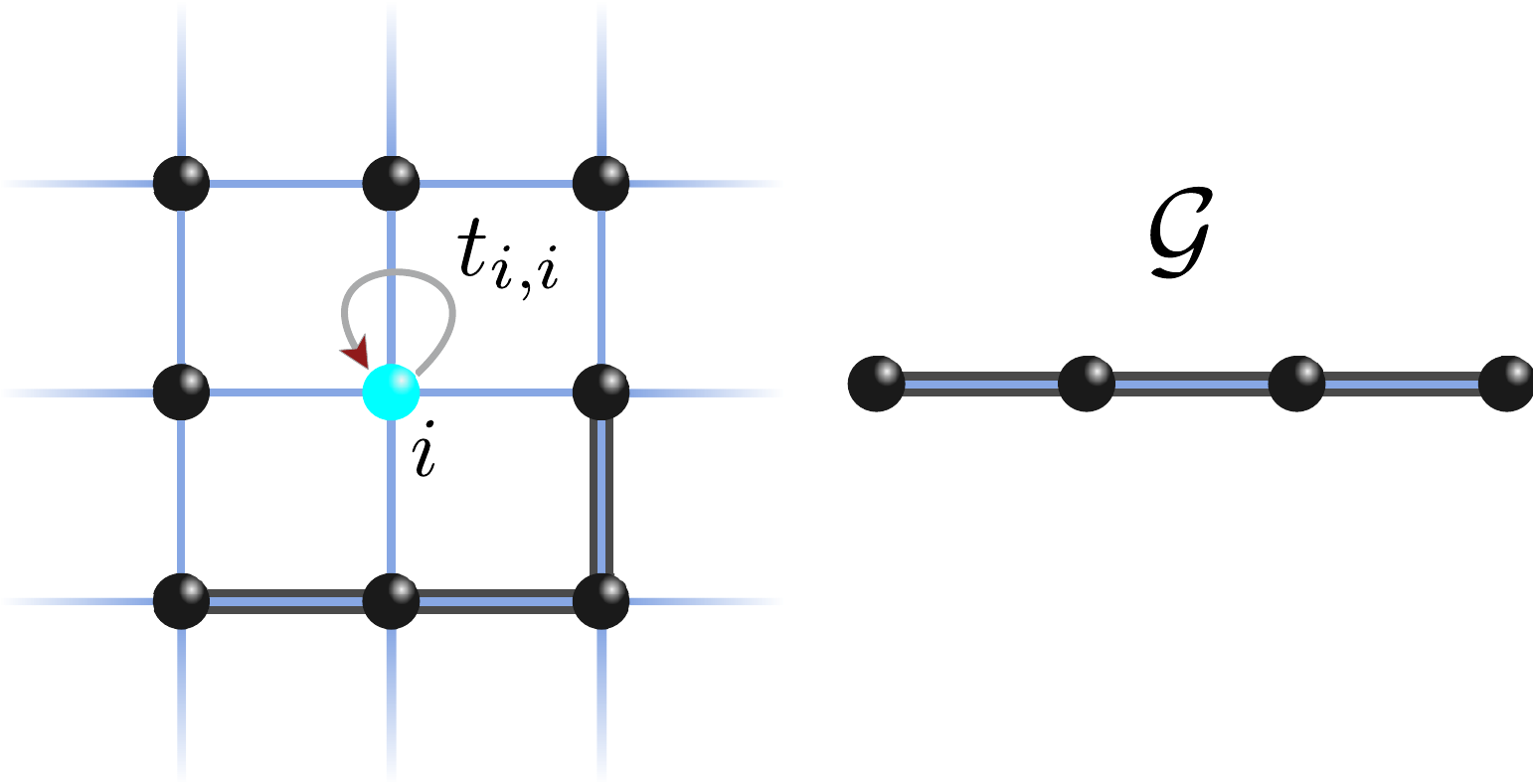}
\end{center}
\caption{The fluctuation pattern indicated by the highlighted links corresponds to fluctuations associated with the vacuum. These fluctuations are irrelevant for the local one-particle hopping element $t_{i,i}$ in the thermodynamic limit (left panel).
} 
\label{fig:groundstate_fluctuations}
\end{figure}

\begin{figure}
\begin{center}
\includegraphics*[width=\columnwidth]{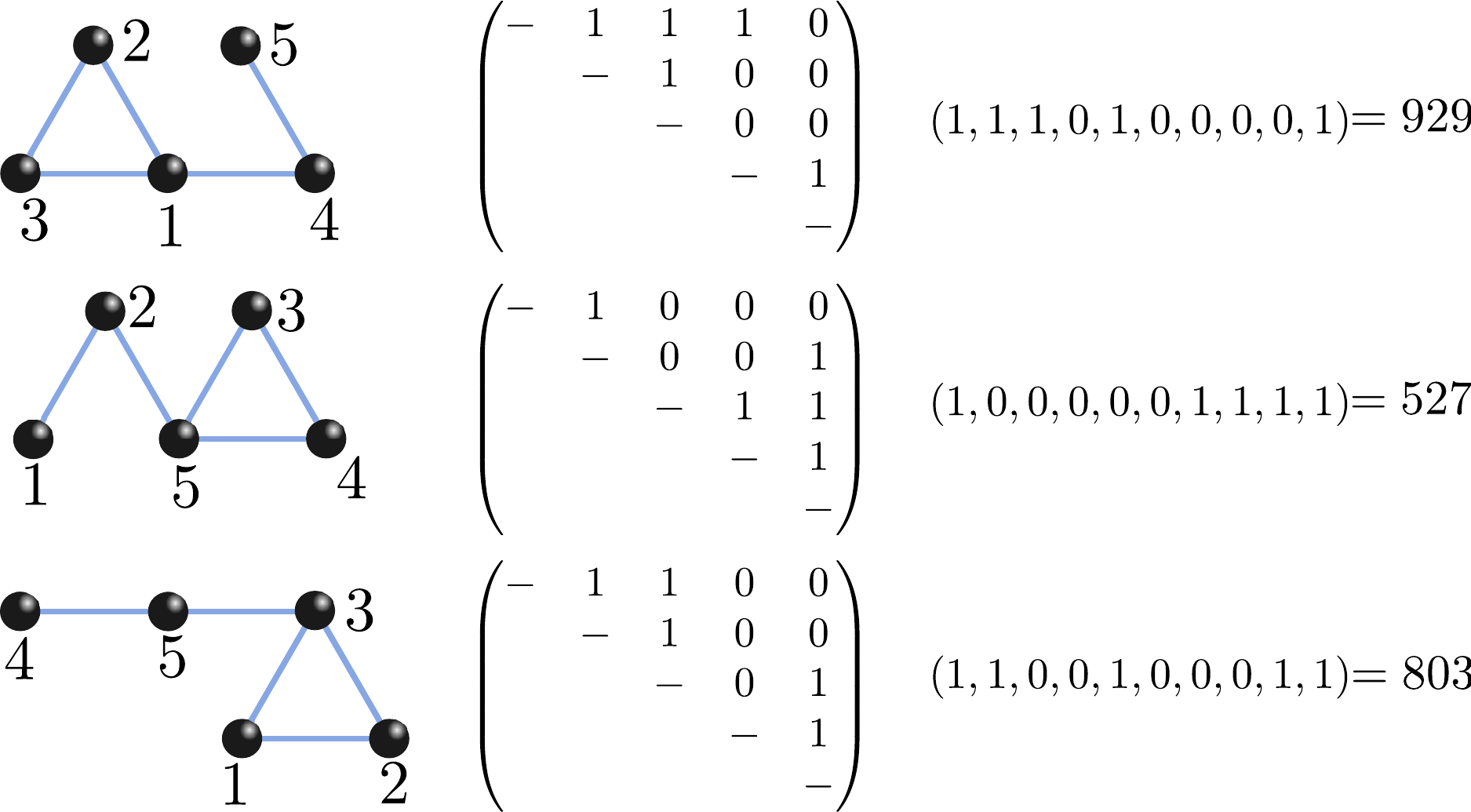}
\end{center}
\caption{Three different labelings of the same graph are depicted with the corresponding adjacent matrices and the resulting keys. The first labeling corresponds to the canonical labeling and the resulting key is the graph key.
} 
\label{fig:different_labelings}
\end{figure}

If the links (and the adjacent sites) form a connected cluster $C$, the process is called linked and the Hamiltonian can be rewritten as sum over these linked clusters 
\begin{widetext}
\begin{align}
\mathcal{H}_\text{eff}(\{\lambda_j\})=\mathcal{H}_0+\sum_{\sum_j k_j=k}^{\infty} \lambda_1^{k_1}\ldots\lambda_{N_\lambda}^{k_{N_\lambda}}\sum_{|\underline{m}|=k\sum_i m_i=0} \sum_{C_k} C(\underline{m}) \sum_{l_1\dots l_k, l_1\bigcup l_2\dots l_k=C_k} \tau_{m_k,l_1}\dots \tau_{m_k,l_k}\label{linked_cluster_decomposition}.
\end{align}
\end{widetext}
Note that there are also fluctuation patterns which correspond to a lower particle channel since not all excitations are involved in the fluctuation as illustrated in Fig.~\ref{fig:different_labelings}. These fluctuations cancel via the subtraction scheme required for the normal ordering.

A linked cluster expansion therefore naturally arises from the pCUT approach. The simplest approach is now to design a sufficiently large cluster which incorporates all the relevant linked fluctuations of a desired order and to carry out the calculation on this single cluster. 
Yet, the calculation based on this approach still incorporates a lot of unlinked processes resulting in an unnecessary computational overhead. Furthermore, these processes are kept in the memory during the calculation which usually defines the highest possible order. 

It is therefore typically favourable to use a full cluster decomposition. To this end the contribution of (small) clusters is calculated by applying $\mathcal{H}_{\text{eff}}$ to all clusters determining the corresponding matrix elements which requires a minimal memory usage. Contributions of smaller subclusters have then to be subtracted in order to gain the reduced contributions of the cluster and to avoid double counting of fluctuations. Hence, after the subtraction, every link $l_j$ in Eq.~\eqref{linked_cluster_decomposition} is involved at least once.

The matrix elements in the thermodynamic limit are then determined by summing up all the reduced contributions of these subclusters. The efficiency of this approach is based on the identification of clusters by means of the topology which implies that the calculation can be restricted to a much smaller set of toplogically distinct clusters called graphs.

\subsubsection{Generating relevant graphs}
In the following, we briefly outline the underlying concepts of the toplogical identification by means of a canonical labeling. The identification process is oriented on the details given in Ref.~\onlinecite{Oitmaa2006}.

First, we consider a system where all (super) sites are coupled by identical undirected bonds. In this case, a cluster is fully determined by an (arbitrary) numeration of the sites and the bonds linking the sites. A cluster of $n$ sites can be represented by a $n\times n$ adjacency matrix $m_{i,j}$ with
\begin{align}
m_{ij}=
\begin{cases}
  1,  & \text{if sites } i \text{ and } j \text{ are connected by a bond}\\
  0, & \text{else.}\label{graph_number_definiert}
\end{cases}
\end{align}
Two clusters are called topologically equivalent, if a simple renumeration of the sites yields the same adjacency matrix in both
 cases. Obviously, it is sufficient to restrict the calculation to topologically distinct clusters (graphs) because the results can easily be mapped via a relabeling.
 
In order to identify clusters as topologically equivalent, it is reasonable to introduce a canonical labeling which is distinct from other labelings. To this end, one identifies the off-diagonal elements $m_{i,j}$ ($i<j$) as the binary bits of an integer number with the order $(m_{1,2},m_{1,3},\dots m_{1,n}\dots m_{n-1,n})$ serving as a key. The labeling which maximizes this key corresponds to the canoncial labeling and defines the graph, see also Fig.~\ref{fig:different_labelings} for examples. If two clusters are topologically equivalent, they are assigned to the same graph. In practice, it is necessary to find the labeling which maximizes the key efficiently. We do not go into technical details for the implementation. For further reading  we refer to Ref.~\onlinecite{Oitmaa2006}.

Using the graph key, it is possible to schematically produce the required connected graphs by successively adding links to a given set of graphs. Furthermore, the number of graphs relevant for the calculation can be additionally reduced. Obviously, it is sufficient to include only graphs that can be embedded into the lattice under consideration, i.e.~if the lattice contains clusters corresponding to this graph.

By construction, the reduced contribution of a graph of $n$ links starts at most at order $n$. Hence, for a given order $n$, all connected graphs of up to $n$ links ahave to be considered. Depending on the problem, further selection rules can help to reduce the number of relevant graphs because the presence of conserved quantities might result in further constraints for a given order. 

\subsubsection{Calculation on graphs}
The next step is to perform the actual calculation on graphs, i.e.~$H_\text{eff}$ is evaluated to determine the relevant low-energy matrix elements. For the implementation, one typically chooses the eigenbasis of $Q$, i.e.~every basis state can be identified with a defined distribution of bare excitations. During the application of the operator sequences in $H_\text{eff}$ the intermediate states are represented as linear combinations of these basis states
\begin{align}\label{eq:intermediate_states}
|\Psi\rangle=\sum_j \alpha_j |j\rangle \quad .
\end{align}
The basis states $|j\rangle$ are represented by integer numbers where the information of the basis states is encoded bitwise in the bit representation of an integer as known for instance from exact diagonalization. In this basis, the action of the operators in Eq.~\eqref{tausumme} affects only a small number of bits allowing for a fast access and modifications via bitwise operations.

Compared to the dimension of the full Hilbert space, the number of terms in an intermediate state is small. Therefore, it is useful to use associative arrays consisting of a key (the state represented by an integer) and an associated value (the coefficient $\alpha_j$ represented by a rational number). This is also illustrated in Fig.~\ref{fig:the_calculation}a.

The desired matrix elements are then given as series expansions with rational coefficients. In order to gain cluster additive quantities for the embedding, the normal ordering is applied followed by the subtraction of subclusters contributions.

\subsubsection{Graph embedding}
Finally, the graphs are embedded into the infinite lattice, i.e.~all realizations of a given graph on the lattice are produced recovering all the clusters in Eq.~\ref{linked_cluster_decomposition} via the appropriate relabeling. The contributions of the cluster additive quantities are added up accordingly. Note that the contributions of all linked clusters in Eq.~\eqref{linked_cluster_decomposition} are incorporated while the actual calculation is performed on a massively smaller subset of topologically distinct clusters. The embedding procedure is straightforward and technical details of the implementation can be found in Ref.~\onlinecite{Oitmaa2006}. 

Typically, the bottleneck of the calculation is defined by the derivation of the effective model on the graphs via pCUTs. Hence, the efficiency of this approach is based on the identifcation of clusters and the associated reduction of the number of clusters involved in the calculations.

\section{White graph expansion}
\label{Sect:White_Graph_Expansion}
Up to now, we have described the essential steps to set up a linked cluster expansion with the pCUT method. This standard scheme becomes inefficient when the physical system possesses many linktypes corresponding to perturbation parameters $\lambda_j$, which is for example relevant when fitting experimental data with microscopic calculations. These linktypes function as another topological attribute which can be incorporated in the usual scheme by generalizing \mbox{$m_{i,j}\in\{0,1\dots n\}$} for $n$ linktypes. Clearly, this leads to an exponential growth of relevant graphs needed for the calculation. However, as we demonstrate in the following, one can circumvent this inconvenience by a so-called white-graph expansion.

The underlying principle of the white-graph expansion is to ignore the different linktypes (colors) for the topological classification of the clusters. The white-graph number is simply defined using Eq.~\eqref{graph_number_definiert} and the calculation is restricted to topologically distinct clusters in this sense. However, the contributions of clusters in Eq.~\eqref{linked_cluster_decomposition} \textit{cannot} be gained by a simple mapping of the sites because the contributions do differ depending on the color pattern of the links. To this end, additional information are tracked during the pCUT calculation on the single graphs allowing to restore the contribution of each cluster exactly from the calculation on white graphs.\\
\subsection{The calculation on white graphs}
\begin{figure}
\begin{center}
\includegraphics*[width=0.95\columnwidth]{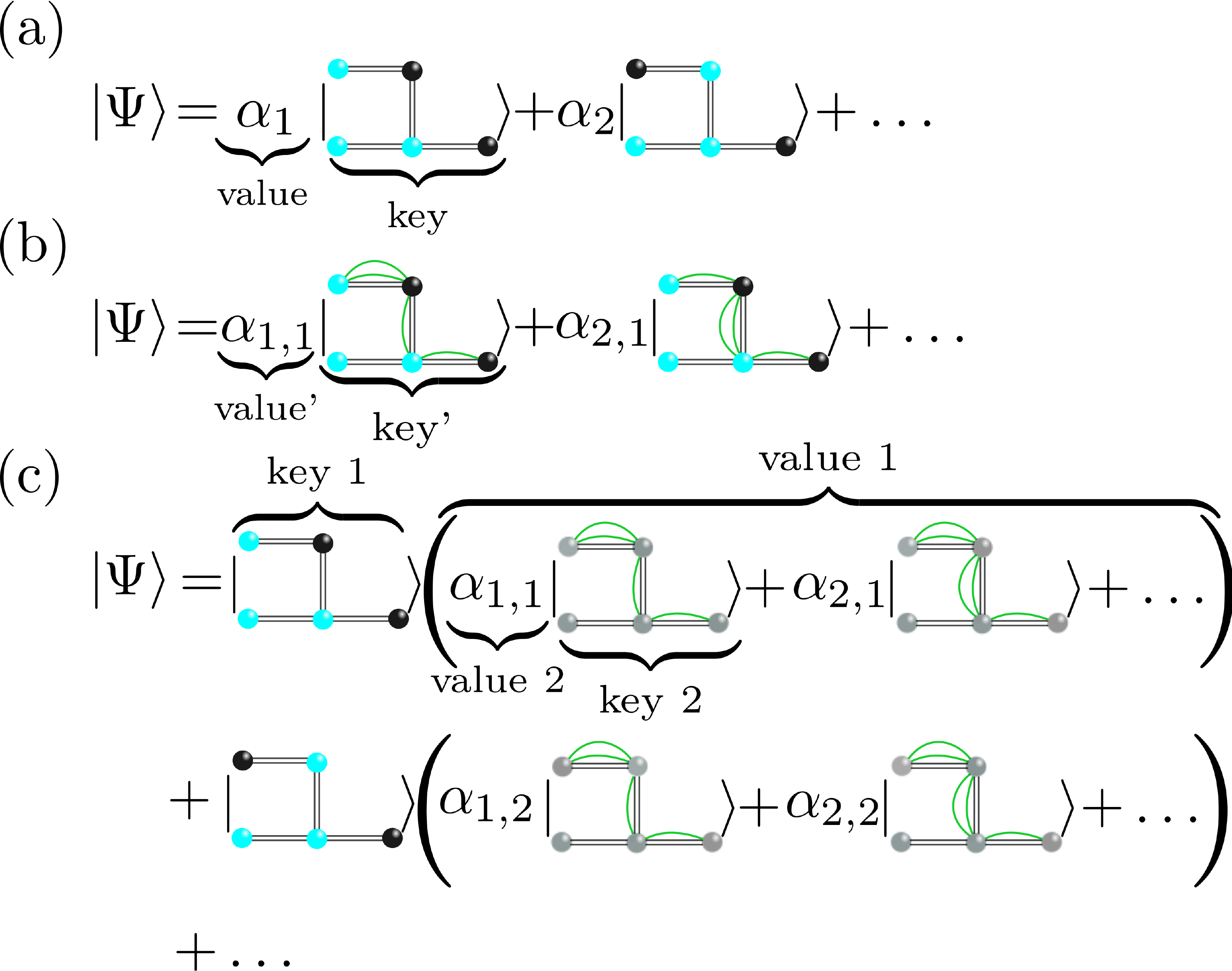}
\end{center}
\caption{(a) The standard representation of intermediate states following Eq.~\eqref{eq:intermediate_states} so that $\alpha_j$ corresponds to a rational number. (b) Standard calculation on white graps expansion using the generalized monoms as written in Eq.~\eqref{eq:white_standard}. (c) Improved calculation on white graphs following Eq.~\eqref{eq:white_improved} which uses the product structure of the generalized monoms.
} 
\label{fig:the_calculation}
\end{figure}
The additional bookkeeping can be interpreted as an enhanced or generalized monom; additional values $\vec{n}$ are assigned to the links to keep track of the relevant information. The latter can be comprised in a single object $M(\vec{n})$ which can be viewed as a generalized monom. The choice of $M(\vec{n})$ is problem dependent and must be adjusted for each application. If the links differ only with respect to the coupling strength, it is sufficient to track how often each link was active in the operator sequence applied to the current state. For more complex problems, more information must be tracked and this approach may become unfavorable in scenarios where a lot of linktypes are involved and the non-zero matrix elements differ greatly between the linktypes. However, in principle it is always possible to track the relevant information.

Evaluated for a given link pattern, the generalized monoms must yield an actual polynom in the multiple perturbation parameters of the different couplings such that the result corresponds exactly to the calculation on a cluster having this very link pattern. This allows for the subsequent evaluation necessary for the embedding procedure.

As indicated by the name, the generalized monoms obey
\begin{align}
M(\vec{n_1})\cdot M(\vec{n_2})=M(\vec{n_1}+\vec{n_2})
\end{align}
and the intermediate states during the calculation are then expressed as linear combinations of these elements:
\begin{align}\label{eq:white_standard}
|\Psi\rangle=\sum_{i,j} \alpha_{i,j} M(\vec{n}_i) |j\rangle \quad .
\end{align}
Note that the action of an operator in Eq.~\eqref{tausumme} affects the generalized monoms $M(\vec{n}_i)$ and the state $|j\rangle$ independently; similarly to the action on product wave functions.

As in a standard pCUT set up, it is reasonable to use associative arrays to represent the intermediate states. The information of a generalized monom can analogously be encoded bitwise. In a naive implementation, the state combined with the generalized monom defines the key and the associated amplitude $\alpha_{i,j}$ corresponds to the value as visualized in Fig.~\ref{fig:the_calculation}b.

In an improved implementation, one can make use of the product structure of the generalized monoms as well as of the intermediate states and factorise the representation:
\begin{align}\label{eq:white_improved}
|\Psi\rangle=\sum_j |j\rangle \left( \sum_i \alpha_{i,j} M(\vec{n}_i)\right) \quad .
\end{align}
The effect of the operators on the state $|j\rangle$ can be calculated independendly of the effect on the appendant sum of generalized monoms. Consequently, numerous redundant search, comparison, and shift operations are avoided.

The implementation of this approach requires two nested containers. The key of the first container is defined by the state $|j\rangle$ while the value is given by a second container. The second container is defined by a key comprising the information of the generalized monom $M(\vec{n}_i)$ and the value corresponds to the amplitude $\alpha_{i,j}$ of this generalized monom and the state. This is illustrated in Fig.~\ref{fig:the_calculation}c.

For an efficient performance, we recommend a simple modification of this procedure. As shown in subsection \ref{ssec:pCUT_LCE}), the reduced contribution of a graph involves each link at least once. This property can help to reduce the computational effort immensely. The objective is to calculate directly the reduced contribution of a cluster without relying on a second subtraction step.

We consider the calculation in order $k$ on a graph consisting of $n$ links. Let $\mu_{i,j}$ denote how often a link $j$ appeared in the operator sequences comprised in the monom $M(\vec{n}_i)$. These information are typically tracked anyway and cause no computational overhead. We define
\begin{align}
\mu_i=\sum_j \tilde{\mu}_{i,j} \quad \text{with }
\begin{cases}
  \tilde{\mu}_{i,j}\equiv\mu_{i,j}-1,  & \text{if } \mu_{i,j}>1\\
  \tilde{\mu}_{i,j}=0, & \text{else.}\label{eta_m}
\end{cases}
\end{align}
One can discard all intermediate generalized monoms $M(\vec{n}_i)$ where $\mu_i>k-n$.\\
Evidently, this is specifically relevant if the number of links $n$ is close or equal to the order $k$, i.e.~for graphs with a large number of links. By construction, these graphs represent the majority of graphs making this modication extremely valueable.

Finally, the matrix elements of the Hamiltonian are given as linear combinations of the generalized monoms. The generalized monoms of these matrix elements build the centerpiece of this approach carrying the relevant information for the embedding of the white graphs described in the following.

\subsection{Embedding white graphs}

For the embedding procedure of a standard linked cluster expansion, all realizations of a given graph on the lattice are produced recovering all the clusters in Eq.~\eqref{linked_cluster_decomposition} via the according relabeling and the contributions of the clusters are added up. Analogously, for the white-graph embedding, all realizations of a given graph on the lattice are generated. In order to recover the contributions of all clusters in Eq.~\ref{linked_cluster_decomposition}, the information of the link pattern must be incorporated in a second evaluation step. The matrix elements on the white graphs are given as sums of generalized monoms. Evaluated for a given link pattern, the generalized monoms yield an polynom in the multiple perturbation parameters recovering the result of the calculation on a cluster corresponding to the given link pattern. The resulting contributions are added up for all embeddings recovering the correct result in the thermodynamic limit.

\section{Coupled two-leg XXZ Heisenberg ladders}
\label{Sect:XXZ_Ladders}
\begin{figure}[t]
\begin{center}
\includegraphics*[width=0.95\columnwidth]{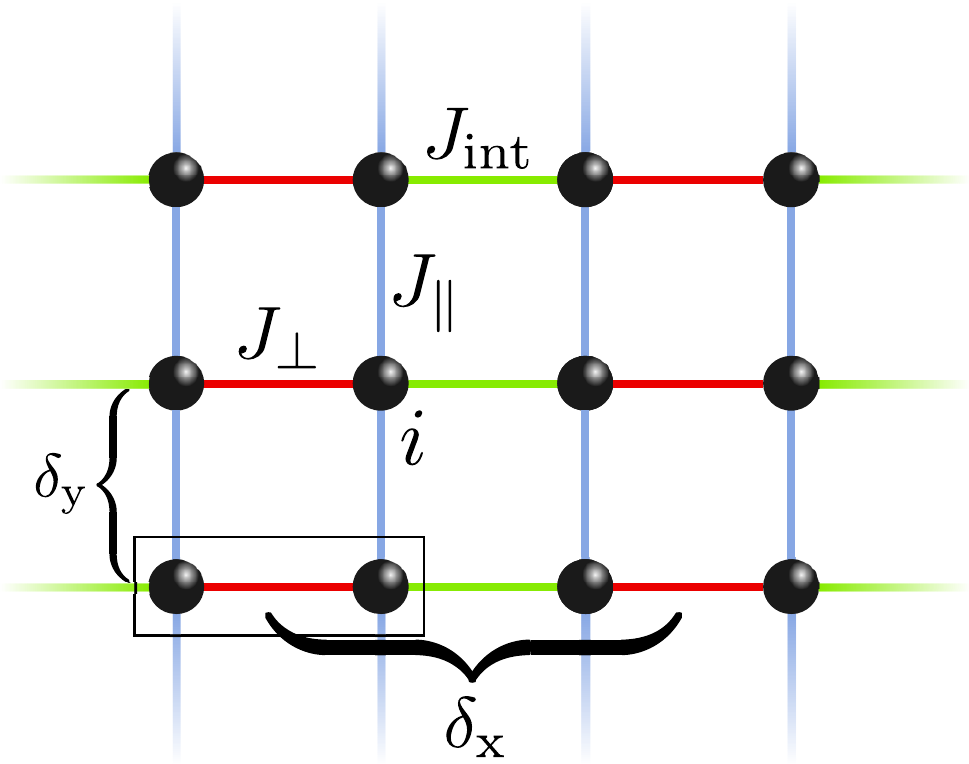}
\end{center}
\caption{Illustration of the quantum spin model of coupled two-leg XXZ ladders. Filled circles represent quantum spin 1/2 which are linked by various solid lines illustrating the different exchange couplings $J_\perp$, $J_\parallel$, and $J_{\rm int}$. Black box shows the two-site unit cell of the system corresponding to the rungs of the ladder and $\delta x$ ($\delta y$) refers to the distance between two neighboring rungs belonging to different (same) ladders.} 
\label{fig:coupled_ladders}
\end{figure}

In order to illustrate the functioning and usefulness of a white-graph expansion, we discuss in
 this section the calculation of one-magnon dispersions for the ordered state of a two-dimensional quantum spin model of coupled two-leg XXZ ladders involving four different pertubation parameters. The microscopic model in terms of spin 1/2 operators reads
\begin{equation} 
\label{ham_unrotated}
 \mathcal{H} =  \sum_{\gamma, \langle i,j\rangle} J_{\gamma} \left[S^{\rm z}_{i} S^{\rm
 z}_{j}+\frac{\lambda}{2}\left(S^{\rm x}_{i}S^{\rm x}_{j}+S^{\rm y}_{i}S^{\rm y}_{j}\right)\right].
\end{equation}
and is illustrated in Fig.~\ref{fig:coupled_ladders}. The different couplings $J_\gamma$ correspond to rung (leg) exchange $J_{\rm rung}$ ($J_{\rm leg}$) of the two-leg ladders, the inter-ladder exchange $J_{\rm int}$, and a spin-anisotropy $\lambda$. This Hamiltonian was recently shown to be relevant for the experimental quantum magnet C$_9$H$_{18}$N$_2$CuBr$_4$ displaying long-range N\'eel order and gapped magnon excitations \cite{Hong2014}.

To calculate the spin-wave dispersion, we perform a sublattice rotation to obtain a ferromagnetic reference state for the Ising case $\lambda=0$, which transforms Eq.~(\ref{ham_unrotated}) to
\begin{equation} \label{ham_rotated}
 \mathcal{H} =  \sum_{\gamma, \langle i,j\rangle} J_{\gamma} \left[-S^{\rm z}_{i} S^{\rm
 z}_{j}-\frac{\lambda}{2}\left(S^{\rm +}_{i}S^{\rm +}_{j}+S^{\rm -}_{i}S^{\rm -}_{j}\right)\right].
\end{equation}
We then introduce hardcore-boson operators $\hat{a}^{\dagger}_\nu$ and $\hat{a}^{\phantom{\dagger}}_\nu$ ($\hat{b}^{\dagger}_\nu$ and $\hat{b}^{\phantom{\dagger}}_\nu$), which create and annihilate a magnon at site $A$ ($B$) of rung $\nu$ above the ferromagnetic reference state, obtaining the hardcore-boson Hamiltonian
\begin{eqnarray}\label{ham_bosons}
 \frac{\mathcal{H}}{\tilde{J}} &=& -\frac{N}{2}+\sum_{\nu}\left( \hat{n}_{\nu}^{(a)}+\hat{n}_{\nu}^{(b)}\right)\nonumber\\
                           && +\sum_\nu \left[ T_{\nu,0}+\lambda\left( T_{\nu,-2}+T_{\nu,+2}\right)\right]\quad ,
\end{eqnarray}
where $\tilde{J}=J_{\rm leg}+(J_{\rm rung}+J_{\rm int})/2$, $N$ is the number of unit cells, $\hat{n}_{\nu}^{(a)}=\hat{a}^\dagger_\nu\hat{a}^{\phantom{\dagger}}_\nu$, and $\hat{n}_{\nu}^{(b)}=\hat{b}^\dagger_\nu\hat{b}^{\phantom{\dagger}}_\nu$. The sums are taken over all rungs. The operators $T_{\nu,n}$, with $T_{\nu,-2}=T_{\nu,+2}^\dagger$, are given by
 \begin{eqnarray}
         T_{\nu,0}    &=& -x_{\rm rung} \hat{n}_{\nu}^{(a)}\hat{n}_{\nu}^{(b)}-x_{\rm int} \hat{n}_{\nu}^{(b)}\hat{n}_{\nu+\delta x}^{(a)}\nonumber\\
                          && -x_{\rm leg} \left( \hat{n}_{\nu}^{(a)}\hat{n}_{\nu+\delta y}^{(a)} + \hat{n}_{\nu}^{(b)}\hat{n}_{\nu+\delta y}^{(b)}\right)
\end{eqnarray}
and
 \begin{eqnarray}
         T_{\nu,+2}    &=& -x_{\rm rung} \hat{a}^\dagger_{\nu}\hat{b}^\dagger_{\nu}-x_{\rm int} \hat{b}^\dagger_{\nu}\hat{a}^\dagger_{\nu+\delta x}\nonumber\\
                           && -x_{\rm leg} \left( \hat{a}^\dagger_{\nu}\hat{a}^\dagger_{\nu+\delta y} + \hat{b}^\dagger_{\nu}\hat{b}^\dagger_{\nu+\delta y}\right) \quad ,
\end{eqnarray}
where $x_{\gamma}=J_{\gamma}/\tilde{J}$, and $\delta x$ ($\delta y$) is the distance between two neighboring rungs belonging to different ladders (the same ladder).

At this point let us replace $x_\gamma\rightarrow\tau x_\gamma$ with \mbox{$\tau\in[0,1]$} so that $\tau=1$ corresponds to our physical system Eq.~\eqref{ham_bosons} and $\tau=0$ functions as a well defined starting point for perturbation theory. Indeed, the resulting Hamiltonian can be rephrased as
\begin{eqnarray}\label{ham_bosons_rephrased}
 \frac{\mathcal{H}}{\tilde{J}} &=& \mathcal{H}_0 + \tau\sum_{n\in\{-2,0,2\}} T_n 
\end{eqnarray}
meeting all criteria relevant to apply the pCUT method as described in detail above: i) The unperturbed Hamiltonian at $\tau=0$ is equidistant, ii) the unperturbed spectrum is bounded from below, and iii) the perturbation is decomposed in $T_n$ operators. One can therefore map Eq.~\eqref{ham_bosons_rephrased} to an effective model, $\mathcal{H}_{\rm eff}$, which conserves the number of magnons. 

The one-magnon sector $\mathcal{H}^{\rm (1)}_{\rm eff}$, which is of our interest here, corresponds in real space to an effective hopping Hamiltonian for the magnons and is therefore fully determined by the one-magnon hopping amplitudes. Here we have calculated all these hopping amplitudes by series expansion up to 13th order in all parameters $\tau x_{\gamma}$. At this order one has to treat $2709$ white graphs in total. Clearly, the number of graphs with color is many orders of magnitude larger and a related calculation is far beyond any realistic set up. 

\begin{figure}[t]
\begin{center}
\includegraphics*[width=0.95\columnwidth]{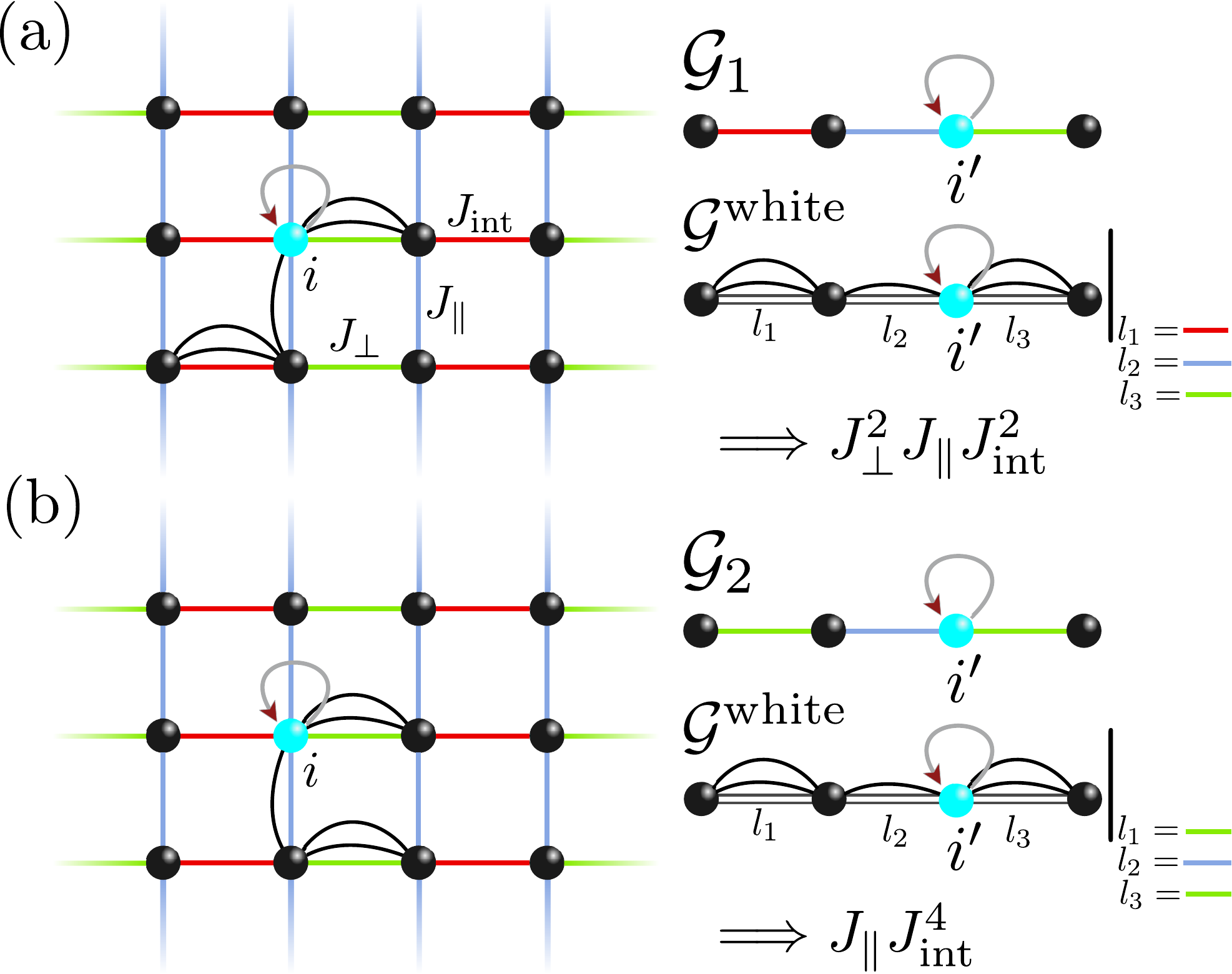}
\end{center}
\caption{Illustration of two different fluctuation patterns contributing to the local one-particle hopping element $t_{i,i}$ in thermodynamic limit (left panel) for the quantum spin model of coupled two-leg XXZ ladders. The solid lines correspond to the different exchange couplings $J_\perp$ (red), $J_\parallel$ (blue), and $J_{\rm int}$ (green). The occurrence of a link in Eq.~\eqref{linked_cluster_decomposition} is visualized by an arc defining the fluctuation pattern. All fluctuations with this pattern are pooled together in this representation. Both fluctuations are associated with topologically distinct graphs $\mathcal{G}_1$ and $\mathcal{G}_2$ (right panel) solely because their colorings are different. Decomposing the results of the calculation with respect to the fluctuation pattern allows to gain both contributions by embedding a single white graph $\mathcal{G}^{\rm white}$ and 'evaluating' the resulting coloring. The evaluation yields polynoms in the different perturbation parameters $J_\gamma$.
} 
\label{fig:white_graph_fluctuations}
\end{figure}

Since the different linktypes differ only with respect to their coupling strengths, the additional information $\vec{n}$ tracked during the calculation on white graphs must incorparate how many times each link is acted on by the perturbation parameters $J_\gamma$. This is in fact identical to the approach where \textit{all} links of a graph have different coupling constants and the result is given as a multi-variable polynom. Finally, during the embedding procedure, the coupling constants of the graph are matched with the ones of the specific graph realizations on the lattice. This is visualized for the local one-particle hopping element $t_{i,i}$ in Fig.~\ref{fig:white_graph_fluctuations}.

The one-magnon sector $\mathcal{H}^{\rm (1)}_{\rm eff}$ can then be simplified by Fourier transform to
\begin{equation}
 \mathcal{H}^{\rm (1)}_{\rm eff} = \sum_{\bm{k}} \left( \omega^{\alpha}(\bm{k}) \alpha^\dagger_{{\bm{k}}} \alpha^{\phantom{\dagger}}_{\bm{k}}+  \omega^{\beta}(\bm{k}) \beta^\dagger_{{\bm{k}}} \beta^{\phantom{\dagger}}_{\bm{k}} \right) \quad ,
\end{equation}
where $\omega^{\alpha}(\bm{k})$ and $\omega^{\beta}(\bm{k})$ denote the two one-magnon branches stemming from the two-site unit cell. 

Let us stress that the results for the one-magnon dispersions are given analytically in all physical parameters $x_\perp$, $x_\parallel$, $x_{\rm int}$, and $\lambda$. Scanning through a large variety of parameter sets, which is often needed when fitting experimental data, is then straighfoward as long as convergence in $\tau$ up to $\tau=1$ is given: the values for the different couplings just have to be inserted into the analytical expression. 

Physically, the series in $\tau$ do converge well for $\lambda$ small. First, both limits $\tau=0$ and $\lambda=0$ have exactly the same ground state as well as the same one-magnon excitations. Only the multi-magnon energies differ. Second, quantum fluctuations are strongly suppressed for small $\lambda$ and/or $\tau$. We find that bare series in $\tau$ are quantitatively converged up to $\tau=1$ for $\lambda\lesssim 0.8$. 

For larger values of $\lambda$, one has to rely on extrapolation schemes like dlogPad\'{e} extrapolation. This is expecially true for the most demanding case of the one-magnon low-energy gap $\Delta\equiv\min_{\vec{k}}\left(\omega^{\alpha}(\vec{k}),\omega^{\beta}(\vec{k})\right)$ at $\lambda=1$. In dlogPad\'{e} extrapolation one constructs various extrapolants $\left[L,M\right]$ where $L$ ($M$) denotes the order of the numerator (denominator). Explicitly, the dlogPad\'e extrapolation is based on the Pad\'{e} extrapolation of the logarithmic derivative of the one-particle gap $\Delta$
\begin{equation}
 \left[\frac{d}{d\tau}\ln \Delta \right]_{[L,M]}:=\frac{P_{L}}{Q_M}\quad ,
 \label{eq:dlog}
\end{equation}
where $P_{L}$ and $Q_M$ are polynomials of order $L$ and $M$. Due to the derivative of the numerator in Eq.~\ref{eq:dlog} one requires $L+M=m-1$ where $m$ denotes the maximum perturbative order which has been calculated (here $m=13$). The $\left[L,M\right]$ dlogPad\'e extrapolant is then given by
\begin{equation}
 \left[L,M\right] :=\exp\left(\int_0^\tau \frac{P_{L}(\tau')}{Q_M(\tau')} d\tau'\right)\quad .
 \label{eq:dlog2}
\end{equation}
In the case of a physical pole at $\tau_0=1$ one is able to determine the dominant power-law behaviour \mbox{$|\tau-\tau_0|^{z\nu}$} close to $\tau_0$. The exponent $z\nu$ is then given by the residuum of $P_L/Q_M$ at $\tau=\tau_0$
\begin{equation}
 z\nu = \frac{P_{L}(\tau)}{\frac{d}{d\tau}Q_M(\tau)} |_{\tau=\tau_0}\quad .
 \label{eq:exp}
\end{equation}
The quality of the dlogPad\'{e} extrapolation can be further improved when the location of a quantum critical point is known exactly. This is true in our case since the only physical point of the expansion is $\tau_0=1$. For this case one can bias the dlogPad\'{e} extrapolation by setting
\begin{equation}
 B\left[L,M\right] :=\exp\left(\int_0^\tau \frac{\bar{P}_{L}(\tau')}{\bar{Q}_M(\tau')}\frac{1}{(\tau-\tau_0)} d\tau'\right)\quad ,
 \label{eq:dlog_bias}
\end{equation}
where 
\begin{equation}
 \left[\left(\tau-\tau_0\right)\frac{d}{d\tau}\ln \Delta \right]_{[L,M]}:=\frac{\bar{P}_{L}}{\bar{Q}_M}\quad .
 \label{eq:dlog_bias2}
\end{equation}
All bias dlogPad\'{e} extrapolants $B\left[L,M\right]$ have a pole at $\tau=\tau_0$ by construction. The value of the denominator of $B\left[L,M\right]$ at $\tau=\tau_0$ corresponds then to the critical exponent $z\nu$. 
 
In the following we exemplify a typical behaviour of $\Delta$ for the isotropic Heisenberg model on the square lattice with $J_{\rm leg}=J_{\rm rung}=J_{\rm int}=\lambda=1$. This isotropic point is known to have gapless Goldstone bosons due to the breaking of SU(2) symmetry in the long-range ordered N\'eel ground state. One therefore expects that the one-magnon gap $\Delta$ is gapped for all values $\tau<1$ and vanishes at $\tau=1$. As outlined in Ref.~\onlinecite{Zheng1991}, the quantum critical behavior is mean-field like, i.e.~one has critical exponents $\nu=1/2$ and $z=1$. 
\begin{figure}[t]
\begin{center}
\includegraphics*[width=0.95\columnwidth]{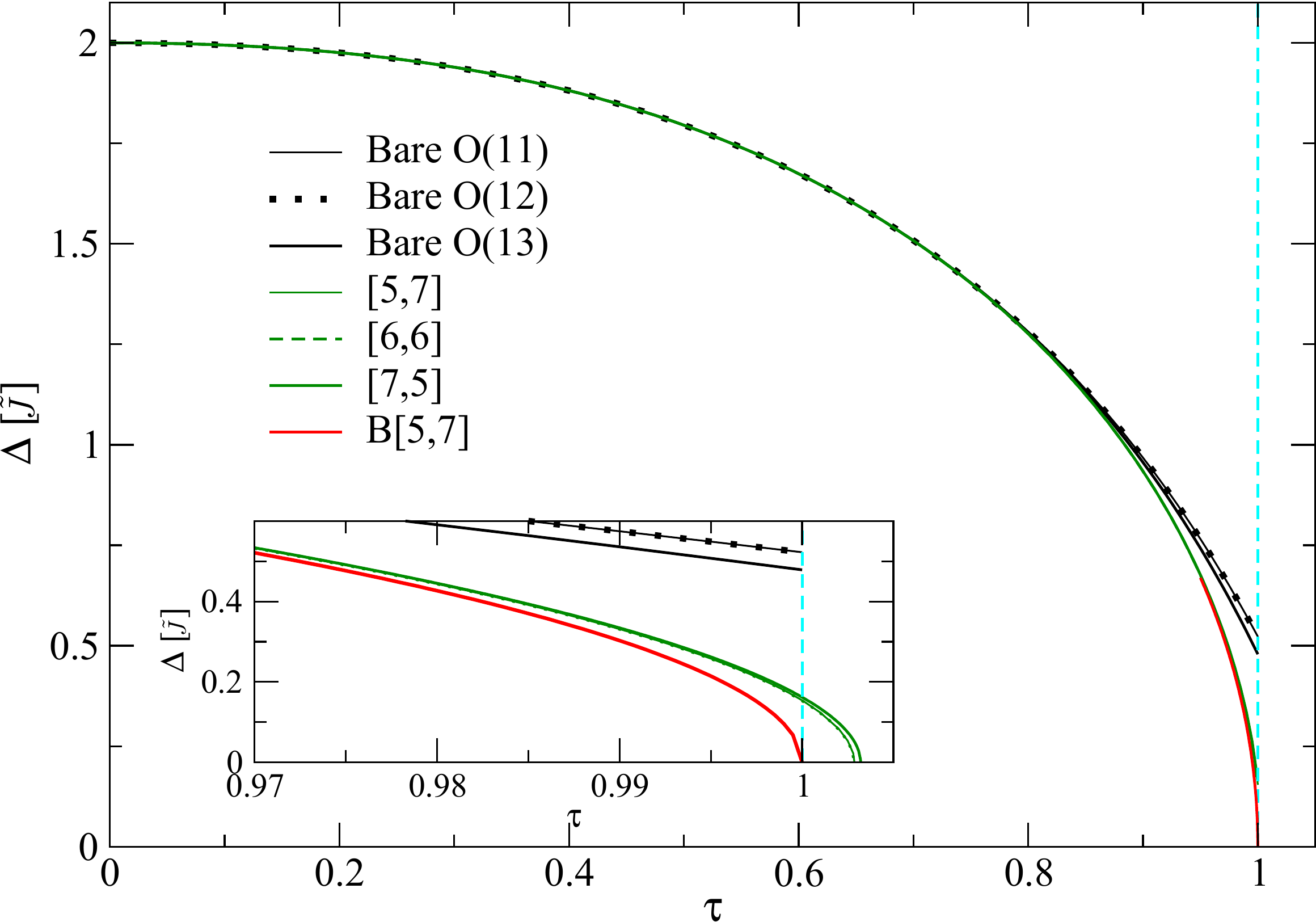}
\end{center}
\caption{One-magnon gap $\Delta/\tilde{J}$ as a function of $\tau$ for the isotropic square lattice $J_{\rm leg}=J_{\rm rung}=J_{\rm int}=\lambda=1$ for different bare series and different (biased) dlogPad\'{e} extrapolants. {\it Inset}: Zoom close to $\tau=1$.} 
\label{fig:gap_vs_tau_isotropic}
\end{figure}
Our results for the one-magnon gap are displayed in Fig.~\ref{fig:gap_vs_tau_isotropic}. The bare series is reliable up to $\tau\lesssim 0.85$ predicting an unphysical gap for $\tau=1$. 

This situation is strongly improved using dlogPad\'{e} extrapolation. Essentially all extrapolants show a pole close to $\tau\approx 1$. Taking all $[L,M]$ of highest order $L+M=12$ with $L\geq 3$ and $L\geq 3$, the average pole is found at $\tau_0=1.004$ with an average critical exponent $z\nu\approx 0.52$ very close to the expected value $1/2$. The quality of the extrapolation gets even better when biasing the extrapolants such that they exhibit a pole at exactly $\tau=1$. In this case one finds an average critical exponent $z\nu\approx 0.499$. Overall, we therefore find quantitative agreement even in the most demanding quantum critical regime using our high-order series expansion. Let us stress that similarly well converged results can be directly deduced in all parts of the (large) parameter space.

\section{Conclusions}
\label{Sect:Conclusions}

In this work we have introduced the so-called white graph expansion for the method of perturbative continuous unitary transformations when implemented
 as a linked cluster expansion. The essential idea behind this expansion is an optimal bookkeeping during the calculation on graphs which is possible due to the presence of the model-independent effective pCUT Hamiltonian in second quantization. Our approach is especially useful when the microscopic model under consideration has several expansion parameters. This case usually represents a major challenge for any kind of linked cluster expansion due to the proliferating number of linked graphs. Indeed, each expansion parameter corresponds to a distinct color in the graph expansion. The white graph expansion overcomes this challenge to a great extent, since the actual calculation is performed on white graphs and the coloring is done after the calculation as a final step. 

We are strongly convinced that this white graph expansion is useful in many situations, e.g.~microscopic models in three dimension or systems with long-rang interactions being notoriously complicated for linked-cluster expansions. Finally, it would be interesting whether other types of linked-cluster expansions like for expample Takahashi's degenerate perturbation theory \cite{Takahashi1977} or matrix perturbation theory \cite{Oitmaa2006} can be also reformulated as a white graph expansion. 

%%%%%%%%%%%%%%%%%%%%%%%%%%%%%%%%%%%%%%%%%%%%%%%%%%%%%%%%%%%%%%%%%%%%%%%%%%%%%%%%
% Acknowledgement
%%%%%%%%%%%%%%%%%%%%%%%%%%%%%%%%%%%%%%%%%%%%%%%%%%%%%%%%%%%%%%%%%%%%%%%%%%%%%%%%
\acknowledgments
This work was in part supported by the Helmholtz Virtual Institute 
``New states of matter and their excitations'' as well as from the Deutsche Forschungsgemeinschaft (DFG) with grant SCHM 2511/9-1. 


%\bibliography{referenzen.bib}
%\bibliographystyle{acm.bst}
%%%%%%%%%%%%%%%%%%%%%%%%%%%%%%%
\end{document}